\pdfoutput=1
\documentclass[conference]{IEEEtran}
\IEEEoverridecommandlockouts

\usepackage{cite}
\usepackage{amsmath,amssymb,amsfonts}
\usepackage{algorithmic}
\usepackage{graphicx}
\usepackage[caption=false,font=footnotesize]{subfig}
\usepackage{textcomp}
\usepackage{xcolor}
\usepackage{booktabs}
\usepackage{multirow}
\usepackage{tikz}
\usetikzlibrary{arrows.meta,decorations.pathmorphing,calc}
\def\BibTeX{{\rm B\kern-.05em{\sc i\kern-.025em b}\kern-.08em
    T\kern-.1667em\lower.7ex\hbox{E}\kern-.125emX}}

\makeatletter
\newcommand{\linebreakand}{%
  \end{@IEEEauthorhalign}
  \hfill\mbox{}\par
  \mbox{}\hfill\begin{@IEEEauthorhalign}}
\makeatother

\newcommand{\bx}{\mathbf{x}}
\newcommand{\bR}{\mathbf{R}}
\newcommand{\bW}{\mathbf{W}}
\newcommand{\bS}{\mathbf{S}}
\newcommand{\bZ}{\mathbf{Z}}
\newcommand{\bI}{\mathbf{I}}
\newcommand{\bu}{\mathbf{u}}
\newcommand{\bv}{\mathbf{v}}
\newcommand{\bh}{\mathbf{h}}
\newcommand{\bA}{\mathbf{A}}
\newcommand{\tr}{\operatorname{tr}}
\newcommand{\ST}{\bS_{\mathrm{TT}}}
\newcommand{\SR}{\bS_{\mathrm{RT}}}

\begin{document}

\title{Electromagnetically Consistent Monostatic ISAC Beamforming: Mutual Coupling, Active Mismatch and Tx--Rx Leakage in Compact Dipole Arrays
}

\author{\IEEEauthorblockN{Stylianos D. Asimonis}
\IEEEauthorblockA{\textit{Dept. of Electrical \& Computer Eng.}\\
\textit{University of Patras}\\
26504 Patras, Greece\\
s.asimonis@upatras.gr}
\and
\IEEEauthorblockN{Christos Tselios}
\IEEEauthorblockA{\textit{Enebula BV} \\
1076DE Amsterdam, Netherlands \\
ctselios@enebula.nl}
\and
\IEEEauthorblockN{Dimitrios Tyrovolas}
\IEEEauthorblockA{\textit{Dept. of Informatics \& Telecom.}\\
\textit{University of Peloponnese}\\
22131 Tripoli, Greece\\
d.tyrovolas@uop.gr}
\linebreakand
\IEEEauthorblockN{Ilias Politis}
\IEEEauthorblockA{\textit{Industrial Systems Institute} \\
\textit{Research Centre ATHENA}\\
26504  Patras, Greece\\
politis@athenarc.gr}
\and
\IEEEauthorblockN{George K. Karagiannidis}
\IEEEauthorblockA{\textit{Dept. of Electrical \& Computer Eng.}\\
\textit{Aristotle University of Thessaloniki}\\
54124 Thessaloniki, Greece\\
geokarag@auth.gr}
}

\maketitle

\begin{abstract}
Beamforming for integrated sensing and communication (ISAC) is usually designed on ideal array manifolds with a norm constraint on the transmit signal. In a compact monostatic transceiver this hides three antenna effects: distorted embedded patterns, beam-dependent active mismatch at the power amplifiers, and transmit-to-receive leakage that raises the radar noise floor. We build a multiport scattering model of a shared-aperture dipole array directly from method-of-moments simulations and use it to formulate a Cramer-Rao bound minimizing ISAC beamformer with multiuser SINR constraints, per-port active reflection constraints and a leakage budget, solved as a semidefinite program with a one-dimensional budget search. Full-wave evaluation shows that coupling-unaware designs miss their SINR targets by up to 6.2 dB and drive amplifiers to an equivalent VSWR above 3.3 at small spacings. For spacings from 0.3 to 0.5 wavelengths, the proposed design meets all targets with an equivalent VSWR below 2 and lowers the root Cramer-Rao bound by 32 to 53 percent for 80 dB of self-interference cancellation and by a factor of 3.4 to 5.6 for 70 dB. An independent finite-difference time-domain model of the final layout in openEMS confirms that the leakage null and the ranking of the designs are preserved.
\end{abstract}


\section{Introduction}
Integrated sensing and communication (ISAC) is a defining feature of 6G, where a base station uses the same spectrum, waveform and hardware to serve users and to sense its surroundings \cite{Liu2020TCOM,Liu2022JSAC}. Multi-antenna transmit beamforming is the main tool for trading off the two functions, and a mature literature designs it to satisfy communication SINR constraints while shaping a radar beampattern \cite{Liu2020TSP} or minimizing the Cram\'er--Rao bound (CRB) of target parameters \cite{Liu2022TSP}.

Almost all of these designs share an electromagnetic abstraction: each antenna is an isotropic point with an ideal steering phase, the antennas do not interact, and the transmit power is $\|\bx\|^2$. For half-wavelength arrays this is a reasonable first approximation. Three trends make it fragile. First, compact and dense arrays are attractive for FR3 and handheld or vehicular ISAC nodes, and mutual coupling (MC) grows rapidly when the spacing drops below $\lambda/2$ \cite{Wallace2004,Ivrlac2010}. Second, a coupled array is a multiport network: the impedance seen by each power amplifier (PA) depends on the excitations of all other ports \cite{Pozar1994}, so a beamformer that ignores this can push PAs into strong mismatch. Third, a monostatic ISAC transceiver receives while it transmits, and the Tx--Rx leakage, which in a compact shared aperture is dominated by near-field coupling between neighbouring elements, must be suppressed by $100$~dB or more to reach the noise floor \cite{Sabharwal2014,Han2025}.

Recent works have started to bring MC into ISAC. Robust beamforming against residual MC uncertainty in patch arrays is studied in \cite{Maeng2026}, and MC in reconfigurable holographic surfaces is modelled with coupled dipole approximations to suppress sidelobes in \cite{Zeng2026}. In metasurface antennas, strong MC has been shown to increase the control over the radiation pattern \cite{Prodhomme2024}. These works focus on the radiated pattern. To the best of our knowledge, the joint effect of MC on the \emph{radiated} pattern, on the \emph{active mismatch} presented to the PAs and on the \emph{Tx--Rx leakage} that sets the sensing noise floor has not been addressed in ISAC beamforming, nor validated with full-wave simulation.

This paper closes that gap with a hardware-grounded approach. Our contributions are:
\begin{itemize}
\item We build a port-level scattering model of a shared-aperture Tx/Rx dipole array directly from NEC-2 method-of-moments simulations \cite{NEC}, giving realized embedded patterns, active reflections and Tx--Rx leakage as linear functions of the PA waves. We validate it against direct full-wave runs.
\item We formulate a CRB-minimizing multiuser ISAC beamformer that, in addition to SINR and power constraints, limits the per-port active reflection (VSWR) and the Tx--Rx leakage, and we link the leakage to the radar noise floor through a self-interference cancellation (SIC) budget. The problem is a semidefinite program (SDP) with an outer one-dimensional search.
\item Full-wave evaluation over element spacings $0.2\lambda$ to $0.5\lambda$ quantifies the SINR loss, PA mismatch and sensing loss of coupling-unaware and pattern-only coupling-aware designs, and the gains of the proposed design.
\item We rebuild the final layout as strip dipoles in an independent FDTD solver (openEMS) and show that the leakage null and the ranking of the designs survive the change of solver and feed model, while the active reflection is the most model-sensitive quantity.
\end{itemize}

\textit{Notation:} Bold lower (upper) case letters denote vectors (matrices); $(\cdot)^T$, $(\cdot)^H$ and $(\cdot)^*$ denote transpose, Hermitian transpose and conjugate; $[\cdot]_{nn}$ is the $n$-th diagonal entry; $\dot{f}=\partial f/\partial\theta$.

\section{Electromagnetic System Model}

\begin{figure}[t]
\centering
\begin{tikzpicture}[scale=0.92, every node/.style={font=\scriptsize}]
\foreach \i in {0,...,7} { \draw[very thick, blue!70!black] (0.42*\i,0) -- (0.42*\i,0.55); \fill[blue!70!black] (0.42*\i,0.275) circle (0.035);}
\foreach \i in {8,...,15} { \draw[very thick, orange!85!black] (0.42*\i,0) -- (0.42*\i,0.55); \fill[orange!85!black] (0.42*\i,0.275) circle (0.035);}
\draw[<->] (0,-0.14) -- (0.42,-0.14); \node at (0.21,-0.32) {$d$};
\node at (1.6,-0.62) {Tx subarray ($N_\mathrm{T}$ PAs)};
\node at (4.83,-0.62) {Rx subarray ($N_\mathrm{R}$ LNAs)};
\draw[-{Latex}, red!70!black, thick, decorate, decoration={snake,amplitude=0.5mm,segment length=1.6mm,post length=1mm}] (2.94,0.7) -- (3.78,0.7);
\node[red!70!black] at (3.36,0.95) {$\SR\bx$};
\node[red!70!black] at (3.36,-0.3) {leakage};
\draw[-{Latex}, thick, gray] (1.3,0.7) -- (0.2,1.95) node[above] {user $1$};
\draw[-{Latex}, thick, gray] (2.3,0.7) -- (3.0,1.95) node[above] {user $K$};
\draw[-{Latex}, thick, black] (1.75,0.7) -- (1.75,2.05) node[above] {target};
\draw[-{Latex}, thick, black, dashed] (1.95,2.05) -- (5.6,0.72) node[pos=0.72, above right=-1pt] {echo};
\node[align=left, font=\scriptsize] at (6.0,1.95) {PA waves $\bx$\\[-1pt]reflections $\ST\bx$\\[-1pt]$|\bar\Gamma_n|\le\Gamma_{\max}$};
\end{tikzpicture}
\caption{Monostatic ISAC transceiver with a shared-aperture collinear array of $z$-directed dipoles. The Tx subarray serves $K$ users and illuminates a target; the Rx subarray receives the echo together with near-field Tx--Rx leakage. Azimuth $\phi$ is measured from the array ($x$) axis.}
\label{fig:sys}
\end{figure}
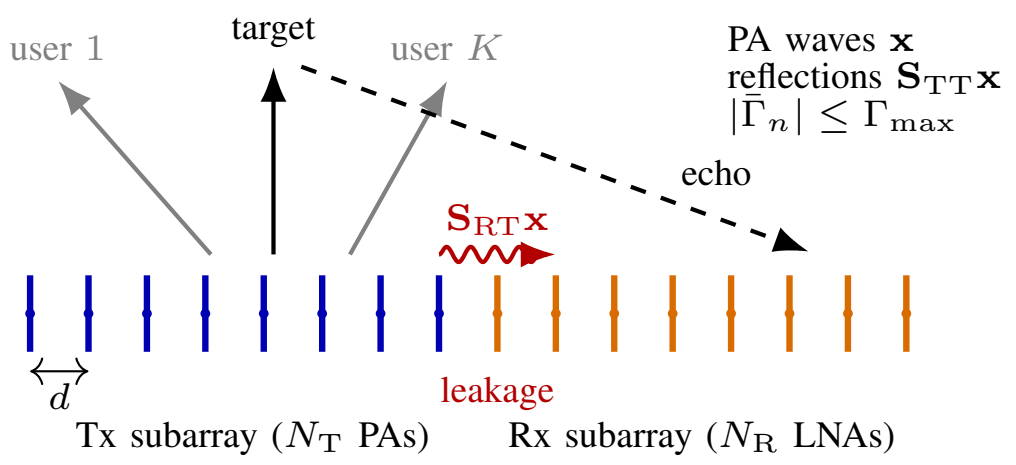

\subsection{Multiport Model from Full-Wave Simulation}
Consider the transceiver of Fig.~\ref{fig:sys}: $N=N_\mathrm{T}+N_\mathrm{R}$ identical $z$-directed thin-wire dipoles, centre-fed, placed along the $x$ axis with spacing $d$. The first $N_\mathrm{T}$ ports are driven by PAs and the last $N_\mathrm{R}$ ports feed low-noise amplifiers (LNAs); all are referenced to $Z_0$. We focus on the azimuth plane $\theta_{\mathrm{el}}=90^\circ$ and denote azimuth by $\phi$, measured from the array axis.

We characterize the array by $N$ method-of-moments runs in NEC-2 \cite{NEC}. In run $m$, port $m$ is driven by a $1$~V gap source and all other ports are short-circuited. The feed currents of all ports give the $m$-th column of the short-circuit admittance matrix $\mathbf{Y}$, and the far field gives the $m$-th column of $\mathbf{G}_{\mathrm{sc}}(\phi)\in\mathbb{C}^{1\times N}$, the field per unit port voltage. By linearity, for any port state, $\bZ=\mathbf{Y}^{-1}$, $\mathbf{v}=\bZ\mathbf{i}$, and the far field is $\mathbf{G}_{\mathrm{sc}}(\phi)\mathbf{v}$. With the power-wave definitions, the scattering matrix is
\begin{equation}
\bS=(\bZ-Z_0\bI)(\bZ+Z_0\bI)^{-1}=\begin{bmatrix}\ST & \bS_{\mathrm{TR}}\\ \SR & \bS_{\mathrm{RR}}\end{bmatrix}.
\end{equation}
Let $\bx\in\mathbb{C}^{N_\mathrm{T}}$ collect the forward waves launched by the PAs, scaled so that $\|\bx\|^2$ is the incident power, while the Rx ports are matched (zero incident wave). The port currents are then $\mathbf{i}=(\bZ+Z_0\bI)^{-1}[2\sqrt{2Z_0}\,\bx^T,\mathbf{0}^T]^T$, and the radiation intensity $U(\phi)$ satisfies $4\pi U(\phi)=|\bu(\phi)\bx|^2$ with the realized embedded pattern row vector
\begin{equation}
\bu(\phi)=\kappa\,\big[\mathbf{G}_{\mathrm{sc}}(\phi)\bZ(\bZ+Z_0\bI)^{-1}\big]_{:,1:N_\mathrm{T}},\quad \kappa=4\sqrt{\pi Z_0/\eta_0},
\label{eq:u}
\end{equation}
where $\eta_0$ is the free-space impedance. Thus $|\bu(\phi)\bx|^2/\|\bx\|^2$ is the realized gain, which includes coupling, mismatch and absorption in the Rx loads. By reciprocity, the receive manifold $\bv(\phi)\in\mathbb{C}^{N_\mathrm{R}}$ is given by the last $N_\mathrm{R}$ columns of the same expression.

The model exposes three hardware quantities that are linear in $\bx$:
\begin{itemize}
\item \emph{Active reflection.} The wave reflected back into PA $n$ is $[\ST\bx]_n$, so the active reflection coefficient is $\Gamma_n=[\ST\bx]_n/x_n$ \cite{Pozar1994}. It depends on the whole beamformer, not only on the element.
\item \emph{Tx--Rx leakage.} The wave delivered to the LNAs is $\SR\bx$, with power $P_\mathrm{L}=\|\SR\bx\|^2$.
\item \emph{Radiated power.} For lossless wires, $P_{\mathrm{rad}}=\|\bx\|^2-\|\ST\bx\|^2-\|\SR\bx\|^2$, which differs from $\|\bx\|^2$.
\end{itemize}
The \emph{ideal} (coupling-unaware) model used by conventional designs instead takes $\bu_{\mathrm{id}}(\phi)=g_0(\phi)[e^{jkx_1\cos\phi},\dots,e^{jkx_{N_\mathrm{T}}\cos\phi}]$, with $g_0$ the realized pattern of an isolated matched dipole and $x_n$ the element positions, and implicitly assumes $\bS=\mathbf{0}$.

\subsection{Communication and Sensing Signals}
The transmitted waves are $\bx=\sum_{k=1}^K\mathbf{w}_ks_k+\bx_\mathrm{s}$, where $s_k$ are unit-power user symbols and $\bx_\mathrm{s}$ is a dedicated sensing signal with covariance $\bR_\mathrm{s}$. The total covariance is $\bR=\sum_k\bW_k+\bR_\mathrm{s}$, $\bW_k=\mathbf{w}_k\mathbf{w}_k^H$, and $\tr(\bR)$ is the total PA output power. User $k$ at azimuth $\phi_k$ receives over a line-of-sight channel $\bh_k=\sqrt{\beta_k}\,\bu(\phi_k)$, where $\beta_k$ is the path gain, so that
\begin{equation}
\mathrm{SINR}_k=\frac{\bh_k\bW_k\bh_k^H}{\sum_{j\neq k}\bh_k\bW_j\bh_k^H+\bh_k\bR_\mathrm{s}\bh_k^H+\sigma_\mathrm{c}^2}.
\end{equation}

For sensing, a point target at azimuth $\theta$ produces at the Rx ports, over $L$ snapshots,
\begin{equation}
\mathbf{Y}_\mathrm{s}=\alpha\,\bA(\theta)\mathbf{X}+\mathbf{N}_\mathrm{SI}+\mathbf{N},\qquad \bA(\theta)=\bv(\theta)\bu(\theta),
\end{equation}
where $|\alpha|^2=\lambda^2\sigma_\mathrm{RCS}/((4\pi)^3r^4)$ follows the radar equation and $\mathbf{N}$ is thermal noise of power $\sigma^2$ per port. The leaked signal $\SR\bx$ is first reduced by analog and digital SIC by $C$~dB \cite{Sabharwal2014}. We model the residual $\mathbf{N}_\mathrm{SI}$ as white with power $10^{-C/10}P_\mathrm{L}/N_\mathrm{R}$ per port, so that the effective noise power is
\begin{equation}
\sigma_\mathrm{R}^2(\bR)=\sigma^2+10^{-C/10}\,\tr\!\big(\SR\bR\SR^H\big)/N_\mathrm{R}.
\label{eq:sigR}
\end{equation}
For large $L$ and sample covariance $\bR$, the CRB of $\theta$ is \cite{Liu2022TSP}
\begin{equation}
\mathrm{CRB}(\theta)=\frac{\sigma_\mathrm{R}^2(\bR)}{2|\alpha|^2L}\cdot\frac{1}{\tr(\dot{\bA}^H\dot{\bA}\bR)-\frac{|\tr(\dot{\bA}^H\bA\bR)|^2}{\tr(\bA^H\bA\bR)}},
\label{eq:crb}
\end{equation}
with $\dot{\bA}=\dot{\bv}\bu+\bv\dot{\bu}$. In (\ref{eq:crb}), coupling enters through $\bu,\bv$ and their derivatives, and leakage enters through $\sigma_\mathrm{R}^2$.

\section{Coupling-, Mismatch- and Leakage-Aware Beamforming}
We minimize the CRB subject to the per-user SINR target $\gamma$, the PA power budget $P_\mathrm{T}$, a per-port active reflection limit $\Gamma_{\max}$ (VSWR limit) and a Tx--Rx leakage budget $\eta$. Since $\sigma_\mathrm{R}^2$ in (\ref{eq:sigR}) is non-decreasing in $P_\mathrm{L}$, for a fixed $\eta$ we have $\sigma_\mathrm{R}^2\le\sigma^2+10^{-C/10}\eta/N_\mathrm{R}$, and minimizing the resulting upper bound on the CRB amounts to maximizing the Fisher term, i.e., the denominator in (\ref{eq:crb}). Using the Schur complement as in \cite{Liu2022TSP}, the per-$\eta$ problem is
\begin{subequations}
\label{eq:P}
\begin{align}
\max_{t,\{\bW_k\},\bR_\mathrm{s}}\ & t\quad \text{s.t.}\ \bW_k\succeq0,\ \bR_\mathrm{s}\succeq0,\\
& \begin{bmatrix}\tr(\dot{\bA}^H\dot{\bA}\bR)-t & \tr(\dot{\bA}^H\bA\bR)\\ \tr(\bA^H\dot{\bA}\bR) & \tr(\bA^H\bA\bR)\end{bmatrix}\succeq0,\label{eq:c1}\\
& (1+\gamma^{-1})\bh_k\bW_k\bh_k^H\ge\bh_k\bR\bh_k^H+\sigma_\mathrm{c}^2,\ \forall k,\label{eq:c2}\\
& \tr(\bR)\le P_\mathrm{T},\label{eq:c3}\\
& [\ST\bR\ST^H]_{nn}\le\Gamma_{\max}^2[\bR]_{nn},\ n=1,\dots,N_\mathrm{T},\label{eq:c4}\\
& \tr(\SR\bR\SR^H)\le\eta.\label{eq:c5}
\end{align}
\end{subequations}
Constraint (\ref{eq:c4}) bounds the power-averaged active reflection at each PA, $\bar{\Gamma}_n^2=[\ST\bR\ST^H]_{nn}/[\bR]_{nn}$, i.e., the ratio of mean reflected to mean incident power, and is linear in $\bR$; so is (\ref{eq:c5}). When $\mathrm{rank}(\bR)>1$, the instantaneous coefficient $[\ST\bx]_n/x_n$ varies with the data, so $(1+\bar\Gamma_n)/(1-\bar\Gamma_n)$ is an \emph{equivalent} VSWR that sets the mean reflected power seen by the PA, not a bound on the peak VSWR of every symbol; a per-symbol guarantee would require symbol-level constraints. Hence (\ref{eq:P}) is an SDP after dropping $\mathrm{rank}(\bW_k)=1$. As in \cite{Liu2020TSP}, a rank-one solution is recovered from any optimal $\{\bW_k,\bR_\mathrm{s}\}$ by $\mathbf{w}_k=\bW_k\bh_k^H/\sqrt{\bh_k\bW_k\bh_k^H}$ and $\bR_\mathrm{s}\leftarrow\bR-\sum_k\mathbf{w}_k\mathbf{w}_k^H\succeq0$. This leaves $\bR$ unchanged, and therefore (\ref{eq:c1}), (\ref{eq:c3})--(\ref{eq:c5}) and every SINR are preserved.

The budget $\eta$ trades the Fisher information that can be delivered by the Tx subarray against the noise floor at the Rx subarray. A tight $\eta$ forces the Tx beam to place weak fields on the Rx elements, which costs degrees of freedom, while a loose $\eta$ lets residual SI dominate (\ref{eq:sigR}). Since (\ref{eq:P}) is solved in a fraction of a second, we select $\eta$ by a one-dimensional grid search that evaluates the \emph{true} CRB (\ref{eq:crb}), including the actual leakage of each solution. 

\textit{Remarks.} (a) Constraint (\ref{eq:c4}) is scale invariant: replacing $\bR$ by $c\bR$, $c>0$, leaves $\bar\Gamma_n$ unchanged. Hence mismatch cannot be repaired by power backoff; it is a property of the beam direction, not of its power. (b) Since $P_\mathrm{L}=\|\SR\bx\|^2$, leakage-free transmission requires $\bx\in\ker\SR$. In our array the singular values of $\SR$ decay very fast: at $d=0.3\lambda$ they are $-4.5$, $-21.1$, $-41.6$ and $-66.6$~dB for the first four modes, so nulling the two strongest leakage modes costs only two of the eight transmit degrees of freedom; this is why deep leakage budgets remain feasible for $d\ge0.3\lambda$.

\textit{Benchmarks.} We compare with: \textbf{CU}, the coupling-unaware design that solves (\ref{eq:c1})--(\ref{eq:c3}) on $\bu_{\mathrm{id}}$; \textbf{PA}, a pattern-aware design that solves (\ref{eq:c1})--(\ref{eq:c3}) on the true embedded patterns (\ref{eq:u}), which captures the radiation-side MC compensation of recent works; and \textbf{VSWR-only}, which adds (\ref{eq:c4}) but not (\ref{eq:c5}). All designs are evaluated on the same NEC-2-derived model.

\section{Full-Wave Results}

\textit{Setup.} The carrier is $f=10$~GHz (FR3, $\lambda=3$~cm). Each dipole is a thin wire of radius $\lambda/200$ divided into $21$ segments; its length $0.46\lambda$ was tuned with NEC-2 to resonance, $Z_\mathrm{in}=72.9+j1.0~\Omega$, and we set $Z_0=73~\Omega$ so that an isolated element is matched (realized gain $2.13$~dBi). The array has $N_\mathrm{T}=N_\mathrm{R}=8$ and a Tx--Rx gap equal to $d$, and we sweep $d\in[0.2,0.5]\lambda$. Simulations use the NEC-2 engine nec2c with a $0.25^\circ$ azimuth grid, and CRB derivatives are taken by central differences. $K=3$ users are located at $\phi_k\in\{50^\circ,110^\circ,135^\circ\}$ at $200$~m under free-space propagation, with $\gamma=12$~dB. The target is at $\theta=75^\circ$, $60$~m, $\sigma_\mathrm{RCS}=1~\mathrm{m}^2$, and $L=256$. We set $P_\mathrm{T}=30$~dBm, bandwidth $100$~MHz and noise figure $9$~dB, giving $\sigma^2=\sigma_\mathrm{c}^2=-85$~dBm. The equivalent-VSWR limit is $2{:}1$ ($\Gamma_{\max}=1/3$), and all VSWR values reported below are equivalent VSWRs $(1+\bar\Gamma_n)/(1-\bar\Gamma_n)$ and $C\in\{70,80,90\}$~dB. Problem (\ref{eq:P}) is solved with CVXPY/Clarabel, and $\eta$ is searched over $[-45,-10]$~dB relative to $P_\mathrm{T}$ in $2.5$~dB steps.

\textit{Model validation.} The extracted $\bZ$ is reciprocal to within $10^{-5}$. A direct NEC-2 run with all Tx ports driven through $Z_0$ and all Rx ports loaded agrees with the reconstruction to relative errors of $9\times10^{-6}$ (currents) and $8\times10^{-5}$ (far field), and its full-sphere radiated power matches the network value $\|\bx\|^2-\|\bS[\bx^T,\mathbf{0}^T]^T\|^2$ to $3.5\times10^{-4}$. Refining the mesh from $21$ to $41$ segments changes $\bZ$ by $4.3\%$ (mainly the feed-gap reactance), but changes the $\sqrt{\mathrm{CRB}}$ of the designs in Table~\ref{tab:metrics} by less than $0.4\%$, their SINR by less than $0.02$~dB and the largest active reflection of the proposed design from $0.333$ to $0.337$.

\begin{figure*}[t]
\centering
\subfloat[]{\includegraphics[width=0.485\textwidth]{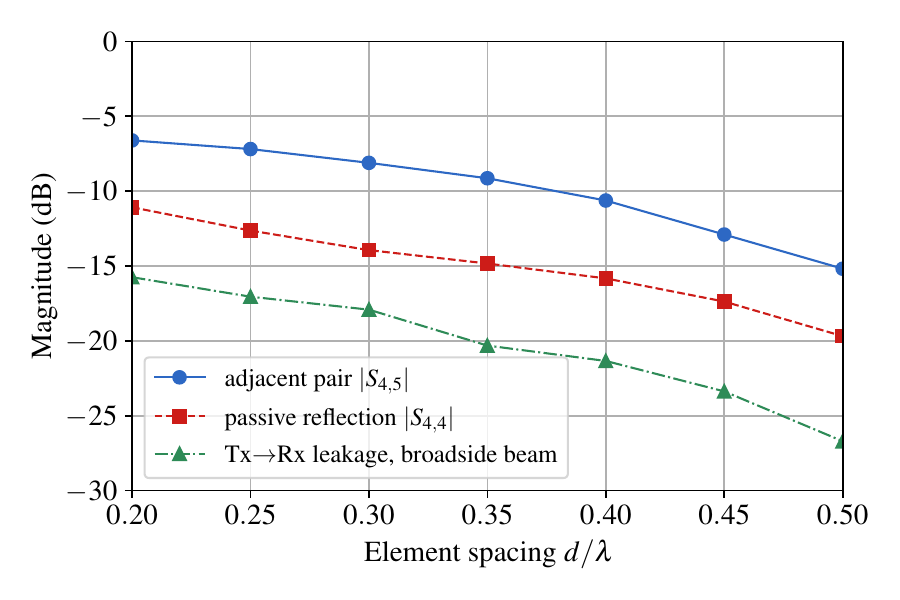}\label{fig:coupling_a}}
\hfil
\subfloat[]{\includegraphics[width=0.485\textwidth]{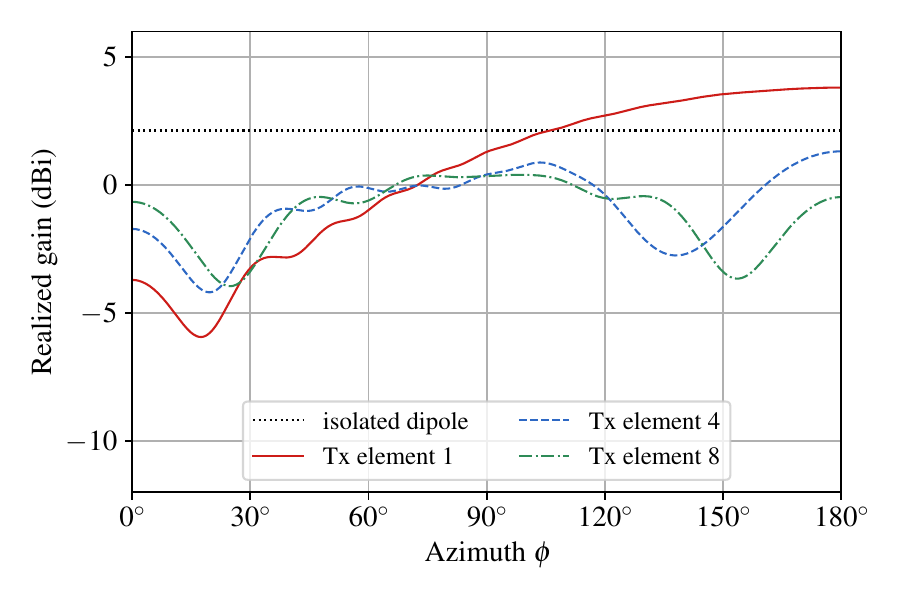}\label{fig:coupling_b}}
\caption{(a)~Coupling between adjacent elements, passive reflection of an inner element and total Tx--Rx leakage of a broadside beam versus spacing. (b)~Realized embedded patterns of Tx elements $1$ (edge), $4$ and $8$ (next to the Rx subarray) at $d=0.3\lambda$, compared with the isolated dipole. Since the collinear array has no structure in $y$, all azimuth patterns satisfy $U(\phi)=U(-\phi)$ and are shown for $0^\circ\le\phi\le180^\circ$.}
\label{fig:coupling}
\end{figure*}

\textit{Coupling.} Fig.~\ref{fig:coupling}(a) shows that the adjacent-element coupling rises from $-15.2$~dB at $0.5\lambda$ to $-8.1$~dB at $0.3\lambda$ and $-6.6$~dB at $0.2\lambda$. The leakage of a simple broadside beam into the Rx subarray is $-17.9$~dB at $0.3\lambda$, so, without beam-domain control, a $30$~dBm transmitter delivers about $12$~dBm to the LNAs. Fig.~\ref{fig:coupling}(b) shows that at $d=0.3\lambda$ the realized embedded patterns deviate from the isolated $2.13$~dBi omnidirectional pattern by several dB and differ from element to element, with the edge element radiating preferentially towards endfire.

\begin{figure}[t]
\centering
\includegraphics[width=\columnwidth]{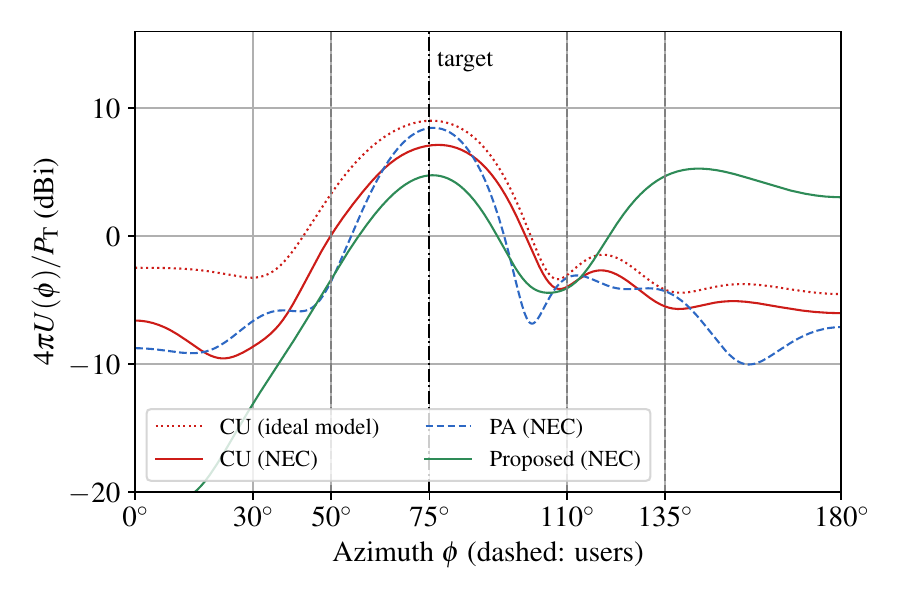}
\caption{Transmit patterns at $d=0.3\lambda$, $C=80$~dB (gain per total PA power), shown for $0^\circ\le\phi\le180^\circ$ by symmetry. Dashed lines: users; dash-dotted line: target.}
\label{fig:pat}
\end{figure}

\textit{Patterns.} Fig.~\ref{fig:pat} compares the patterns at $d=0.3\lambda$. The CU design looks correct on the ideal model (dotted), but on the real array its main beam is squinted by $1.7^\circ$ and its gain towards the target drops by $1.9$~dB. More importantly, its user SINRs fall to $10.9$, $9.0$ and $6.8$~dB instead of $12$~dB. The PA design restores the SINRs and the main beam. The proposed design has a gain towards the target is $3.7$~dB \emph{lower} than PA and it places a strong lobe around $145^\circ$, away from the Rx subarray (which lies towards $\phi=0^\circ$). This beam reduces the Tx--Rx leakage from $-16.5$~dB to $-37.5$~dB relative to $P_\mathrm{T}$, which brings the residual SI per Rx port from $-75.5$~dBm, $9.9$~dB above thermal noise, to $-96.5$~dBm, below thermal noise.

\begin{table}[t]
\caption{Full-wave metrics at $C=80$~dB (target SINR $12$~dB)}
\label{tab:metrics}
\centering
\setlength{\tabcolsep}{2.4pt}
\begin{tabular}{llccccc}
\toprule
$d$ & Design & min SINR & max eq. & $P_\mathrm{L}/P_\mathrm{T}$ & $P_{\mathrm{rad}}/P_\mathrm{T}$ & $\sqrt{\mathrm{CRB}}$\\
 & & (dB) & VSWR & (dB) & (\%) & (deg)\\
\midrule
\multirow{4}{*}{$0.3\lambda$} & CU & $6.8$ & $3.36$ & $-14.8$ & $82.5$ & $2.53$\\
 & PA & $12.0$ & $3.57$ & $-16.5$ & $82.2$ & $2.11$\\
 & VSWR-only & $12.0$ & $2.00$ & $-12.7$ & $83.5$ & $3.81$\\
 & Proposed & $12.0$ & $2.00$ & $-37.5$ & $88.9$ & $\mathbf{1.12}$\\
\midrule
\multirow{4}{*}{$0.4\lambda$} & CU & $9.5$ & $2.41$ & $-19.2$ & $89.2$ & $0.88$\\
 & PA & $12.0$ & $2.44$ & $-19.0$ & $88.8$ & $0.90$\\
 & VSWR-only & $12.0$ & $2.00$ & $-19.4$ & $89.4$ & $0.87$\\
 & Proposed & $12.0$ & $2.00$ & $-35.0$ & $90.2$ & $\mathbf{0.41}$\\
\bottomrule
\end{tabular}
\end{table}

\textit{Hardware metrics.} Table~\ref{tab:metrics} summarizes the full-wave metrics. Both CU and PA drive at least one PA into an equivalent VSWR of $3.4$--$3.6$ at $0.3\lambda$, i.e., on average about $30\%$ of its incident power is reflected back into it. The VSWR-only design fixes the mismatch but, without (\ref{eq:c5}), it moves energy into the leakage path and its CRB is the worst of all. The proposed design satisfies every constraint and also radiates more of the PA power ($88.9\%$ versus $82.2\%$ for PA at $0.3\lambda$), because less power is reflected and less is absorbed by the Rx loads.

\begin{figure}[t]
\centering
\includegraphics[width=\columnwidth]{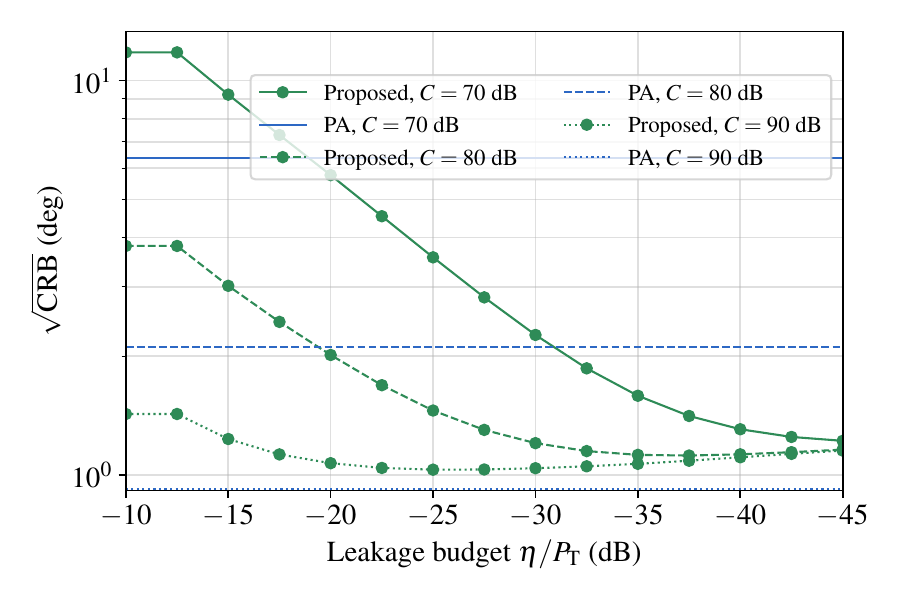}
\caption{Root-CRB of the proposed design versus leakage budget $\eta$ at $d=0.3\lambda$ for three SIC budgets, compared with the PA design.}
\label{fig:eta}
\end{figure}

\textit{Leakage budget.} Fig.~\ref{fig:eta} shows the CRB versus $\eta$. For each SIC budget $C$, the curve has a minimum: tightening $\eta$ first lowers the noise floor quickly and then starts to cost Fisher information. The optimum moves to tighter budgets as $C$ decreases ($-25$~dB for $C=90$~dB, $-37.5$~dB for $C=80$~dB and $-45$~dB, the edge of the grid, for $C=70$~dB). At $C=70$~dB the proposed design achieves $1.22^\circ$ against $6.35^\circ$ for PA, a $5.2\times$ improvement. At $C=90$~dB the receiver is almost thermal-noise limited for all designs, and the proposed design is $12\%$ worse than PA at $d=0.3\lambda$. This is the price of the VSWR constraint, which PA violates; for $d\ge0.35\lambda$ the proposed design is still $2$--$14\%$ better at $C=90$~dB.

\begin{figure}[t]
\centering
\includegraphics[width=\columnwidth]{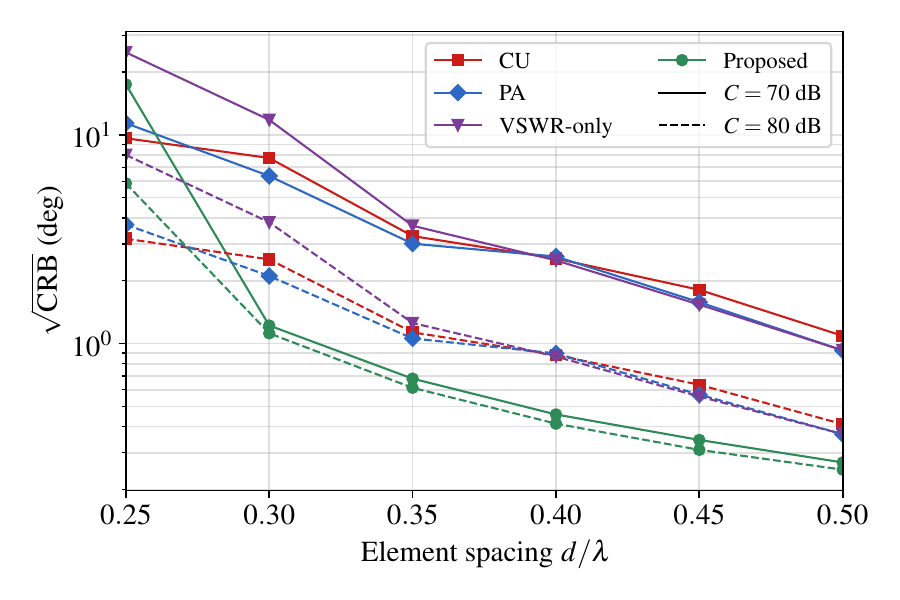}
\caption{Root-CRB versus element spacing for $C=70$~dB (solid) and $C=80$~dB (dashed). Missing points: VSWR-compliant designs infeasible.}
\label{fig:d}
\end{figure}

\begin{figure*}[t]
\centering
\subfloat[]{\includegraphics[width=0.33\textwidth]{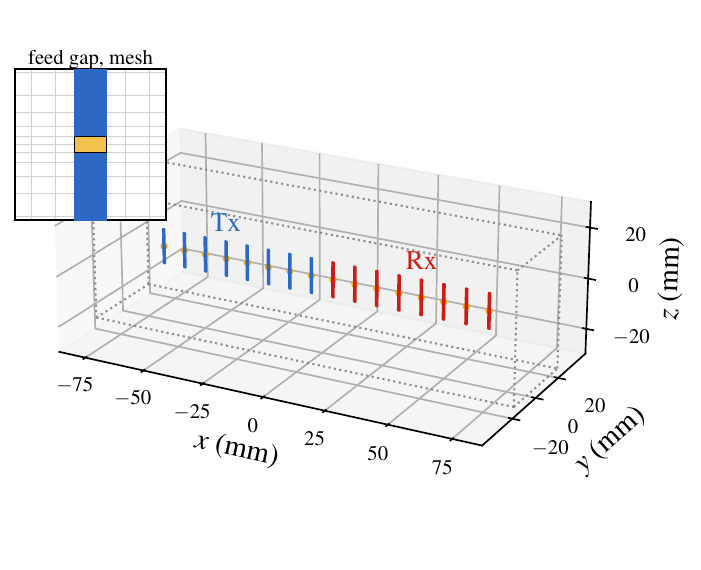}\label{fig:oems_a}}
\hfil
\subfloat[]{\includegraphics[width=0.32\textwidth]{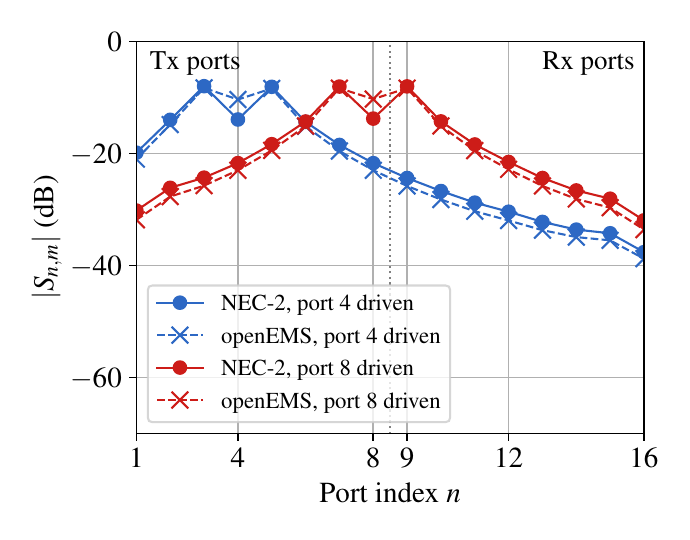}\label{fig:oems_b}}
\hfil
\subfloat[]{\includegraphics[width=0.32\textwidth]{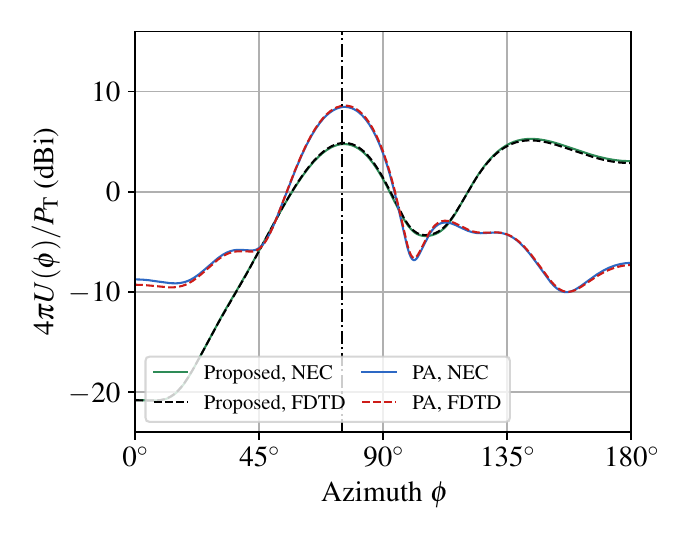}\label{fig:oems_c}}
\caption{Independent openEMS (FDTD) model of the $d=0.3\lambda$ layout. (a)~Simulated geometry: $16$ PEC strip dipoles (Tx in blue, Rx in red) with lumped ports, inside the PML-terminated domain; the inset shows the feed gap and the local mesh. (b)~Scattering-matrix columns for an inner Tx port and for the Tx port next to the Rx subarray. (c)~Transmit patterns ($0^\circ\le\phi\le180^\circ$, by symmetry) of the NEC-designed proposed and PA beamformers on the NEC-2 and FDTD models, without re-optimization.}
\label{fig:oems}
\end{figure*}

\textit{Spacing.} Fig.~\ref{fig:d} shows the CRB versus $d$. For $d\ge0.3\lambda$, the proposed design is consistently the best, reducing the root-CRB of the best benchmark by $32$--$53\%$ at $C=80$~dB and by $3.4$--$5.6\times$ at $C=70$~dB. Even at $d=0.5\lambda$, where coupling is often considered negligible, CU misses the SINR target by $0.7$~dB and the proposed design improves the root-CRB by $32\%$ at $C=80$~dB, because leakage still limits the receiver. At $d=0.2\lambda$, CU misses the target by $6.2$~dB and both CU and PA reach VSWR values of $5.0$--$5.5$. Below $0.3\lambda$ the picture changes: at $0.25\lambda$ the VSWR constraint can be met only with leakage budgets of $-17.5$~dB or looser, and the proposed design is worse than the (non-compliant) benchmarks, while at $0.2\lambda$ no design meets both the SINR targets and the $2{:}1$ VSWR limit. Here compactness must be paid for by matching or decoupling networks.

\section{Independent FDTD Verification of the Final Layout}
To test the conclusions against a different numerical method and feed model, we rebuilt the $d=0.3\lambda$ layout (Fig.~\ref{fig:oems}(a)) in the open-source finite-difference time-domain (FDTD) solver openEMS \cite{openEMS}. Each wire is replaced by a zero-thickness PEC strip of width $w=4a=0.6$~mm, the strip whose equivalent radius equals the NEC wire radius, and each dipole is fed by a lumped $73~\Omega$ port across a $0.3$~mm gap. The computational domain is terminated by $8$-cell perfectly matched layers, the mesh is graded down to $w/2$ at the strips and the gap, and a Gaussian pulse covers $7$--$13$~GHz. Sixteen runs, one per driven port with all other ports terminated in $73~\Omega$, give the full $16\times16$ scattering matrix and, through a near-to-far-field transformation, the realized embedded patterns in the same normalization as (\ref{eq:u}). No quantity is copied from the NEC model.

\begin{table}[t]
\caption{Metrics on the openEMS model ($d=0.3\lambda$, $C=80$~dB)}
\label{tab:oems}
\centering
\setlength{\tabcolsep}{2.0pt}
\begin{tabular}{llccccc}
\toprule
Opt. on & Design & min SINR & max eq. & $P_\mathrm{L}/P_\mathrm{T}$ & $P_{\mathrm{rad}}/P_\mathrm{T}$ & $\sqrt{\mathrm{CRB}}$\\
 & & (dB) & VSWR & (dB) & (\%) & (deg)\\
\midrule
\multirow{4}{*}{NEC} & CU & $6.7$ & $2.69$ & $-15.3$ & $85.5$ & $2.35$\\
 & PA & $11.9$ & $3.03$ & $-16.9$ & $85.4$ & $1.97$\\
 & VSWR-only & $11.7$ & $2.21$ & $-13.2$ & $85.0$ & $3.49$\\
 & Proposed & $12.0$ & $2.18$ & $-37.4$ & $89.0$ & $1.09$\\
\midrule
\multirow{2}{*}{FDTD} & PA & $12.0$ & $2.92$ & $-16.9$ & $85.4$ & $1.97$\\
 & Proposed & $12.0$ & $2.00$ & $-37.5$ & $88.9$ & $\mathbf{0.87}$\\
\bottomrule
\end{tabular}
\end{table}

\textit{Agreement of the two solvers.} The FDTD scattering matrix is reciprocal to within $0.25\%$. Fig.~\ref{fig:oems}(b) shows that the mutual terms agree with NEC-2 over a $35$~dB range (to within $1.7$~dB for all $240$ mutual terms), from $-8.1$ versus $-8.4$~dB for the adjacent elements to $-42.1$ versus $-42.8$~dB for the most distant pair. The self terms differ: the strip dipoles with a finite gap are slightly detuned, and the passive reflection of an inner element rises from $-13.9$~dB (NEC-2) to $-10.4$~dB (FDTD). Over all $16$ realized embedded patterns, wherever the gain exceeds $-5$~dBi, the two solvers differ by $0.28$~dB rms and $0.68$~dB at most.

\textit{Model mismatch.} The upper part of Table~\ref{tab:oems} replays the beamformers designed on the NEC-2 model, without any change, on the FDTD model, which here plays the role of the ``real'' array. The proposed design keeps its SINRs within $0.05$~dB of the target and its leakage at $-37.4$~dB, i.e., the beam-domain leakage null is robust to the change of solver and feed model, because it depends on the mutual terms, which both solvers agree on. Its root-CRB is $1.09^\circ$, against $1.97^\circ$ for PA and $2.35^\circ$ for CU, and Fig.~\ref{fig:oems}(c) shows that the patterns are practically unchanged. The quantity that moves is the active reflection, which depends on the self terms: the largest VSWR of the proposed design rises from $2.00$ to $2.18$, and the PA design still reaches $3.03$. In practice this suggests a small margin on $\Gamma_{\max}$, or re-optimization on the final model.

\begin{figure}[t]
\centering
\includegraphics[width=\columnwidth]{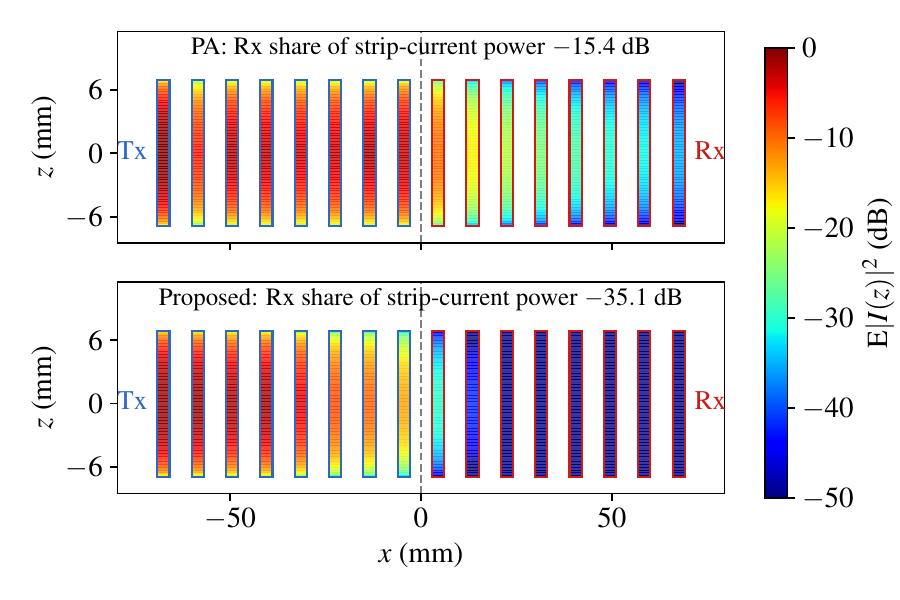}
\caption{FDTD strip currents on the $d=0.3\lambda$ layout for the NEC-designed PA (top) and proposed (bottom) beamformers: $\mathrm{E}|I(z)|^2$ along each dipole, from the superposition of the eight Tx-port runs weighted by $\bR$, on a common normalization (strip width exaggerated for visibility).}
\label{fig:fields}
\end{figure}

\textit{Where the leakage goes.} Fig.~\ref{fig:fields} shows the leakage constraint (\ref{eq:c5}) at work on the physical layout. The strip currents are obtained from the jump of the tangential magnetic field across the strip plane and superposed with the covariance $\bR$ of each beamformer. The PA beamformer drives all Tx dipoles within about $3$~dB of each other, including the one next to the gap, and the first Rx dipole carries a current only $10.2$~dB below the strongest Tx current; in total the Rx strips carry $-15.4$~dB of the strip-current power. The proposed beamformer tapers the Tx currents towards the gap (the eighth dipole is $13$~dB below the first) and phases them so that their fields cancel on the Rx side: the first Rx dipole drops to $-29.8$~dB and the Rx share to $-35.1$~dB. These values agree with the port-level leakage of Table~\ref{tab:oems} ($-16.9$ and $-37.4$~dB), so the leakage null is a near-field cancellation on the actual structure, not an artefact of the network model.

\textit{Re-optimization.} The lower part of Table~\ref{tab:oems} solves (\ref{eq:P}) directly on the FDTD data. The proposed design again meets every constraint, selects the same leakage budget ($-37.5$~dB) and reaches $0.87^\circ$, which is $56\%$ below the PA design optimized on the same data; the latter still drives a PA to a VSWR of $2.92$. The relative ranking of all designs is thus the same in both solvers, which supports the conclusions of Section~IV.

\section{Conclusion}
We showed that, in compact monostatic ISAC transceivers, the antenna array determines not only the radiated beam but also the load seen by the PAs and the noise floor of the radar receiver. A multiport model extracted from NEC-2 makes all three linear in the PA waves, so that VSWR and Tx--Rx leakage constraints fit naturally into a CRB-minimizing ISAC SDP. Full-wave evaluation shows that coupling-unaware designs violate SINR and VSWR requirements, and that controlling leakage in the beam domain can reduce the angle root-CRB by $32$--$53\%$ at $80$~dB SIC and by $3.4$--$5.6\times$ at $70$~dB, for spacings from $0.3\lambda$ to $0.5\lambda$. An independent FDTD model of the final layout confirms that the leakage null is robust to the solver and feed model, whereas the active reflection depends on the feed details and calls for a small design margin. Future work will include dipoles over a ground plane on a substrate, wideband (OFDM) operation, robust designs against fabrication tolerances, and measurements on a prototype.


\begin{thebibliography}{99}
\bibitem{Liu2020TCOM} F.~Liu, C.~Masouros, A.~P.~Petropulu, H.~Griffiths, and L.~Hanzo, ``Joint radar and communication design: Applications, state-of-the-art, and the road ahead,'' \emph{IEEE Trans. Commun.}, vol.~68, no.~6, pp.~3834--3862, Jun. 2020.
\bibitem{Liu2022JSAC} F.~Liu, Y.~Cui, C.~Masouros, J.~Xu, T.~X.~Han, Y.~C.~Eldar, and S.~Buzzi, ``Integrated sensing and communications: Toward dual-functional wireless networks for 6G and beyond,'' \emph{IEEE J. Sel. Areas Commun.}, vol.~40, no.~6, pp.~1728--1767, Jun. 2022.
\bibitem{Liu2020TSP} X.~Liu, T.~Huang, N.~Shlezinger, Y.~Liu, J.~Zhou, and Y.~C.~Eldar, ``Joint transmit beamforming for multiuser MIMO communications and MIMO radar,'' \emph{IEEE Trans. Signal Process.}, vol.~68, pp.~3929--3944, 2020.
\bibitem{Liu2022TSP} F.~Liu, Y.-F.~Liu, A.~Li, C.~Masouros, and Y.~C.~Eldar, ``Cram\'er-Rao bound optimization for joint radar-communication beamforming,'' \emph{IEEE Trans. Signal Process.}, vol.~70, pp.~240--253, 2022.
\bibitem{Wallace2004} J.~W.~Wallace and M.~A.~Jensen, ``Mutual coupling in MIMO wireless systems: A rigorous network theory analysis,'' \emph{IEEE Trans. Wireless Commun.}, vol.~3, no.~4, pp.~1317--1325, Jul. 2004.
\bibitem{Ivrlac2010} M.~T.~Ivrla\v{c} and J.~A.~Nossek, ``Toward a circuit theory of communication,'' \emph{IEEE Trans. Circuits Syst. I}, vol.~57, no.~7, pp.~1663--1683, Jul. 2010.
\bibitem{Pozar1994} D.~M.~Pozar, ``The active element pattern,'' \emph{IEEE Trans. Antennas Propag.}, vol.~42, no.~8, pp.~1176--1178, Aug. 1994.
\bibitem{Sabharwal2014} A.~Sabharwal, P.~Schniter, D.~Guo, D.~W.~Bliss, S.~Rangarajan, and R.~Wichman, ``In-band full-duplex wireless: Challenges and opportunities,'' \emph{IEEE J. Sel. Areas Commun.}, vol.~32, no.~9, pp.~1637--1652, Sep. 2014.
\bibitem{Han2025} K.~Han, C.~Masouros, T.~Riihonen, and M.~G.~Amin, ``Next-generation MIMO transceivers for integrated sensing and communications: Unique security vulnerabilities and solutions,'' arXiv:2511.20309, 2025.
\bibitem{Maeng2026} J.~Maeng and K.~Han, ``Robust beamforming design for integrated sensing and communications with mutual coupling effect,'' arXiv:2608.15556, 2026.
\bibitem{Zeng2026} S.~Zeng, H.~Zhang, B.~Di, H.~Zhang, Z.~Shao, Z.~Han, H.~V.~Poor, and L.~Song, ``Holographic beamforming for integrated sensing and communication with mutual coupling effects,'' \emph{IEEE J. Sel. Areas Commun.}, vol.~44, pp.~480--497, 2026.
\bibitem{Prodhomme2024} H.~Prod'homme and P.~del~Hougne, ``Mutual coupling in dynamic metasurface antennas: Foe, but also friend,'' arXiv:2412.01002, 2024.
\bibitem{NEC} G.~J.~Burke and A.~J.~Poggio, ``Numerical Electromagnetics Code (NEC): Method of moments,'' Lawrence Livermore Lab., Tech. Rep. UCID-18834, 1981.
\bibitem{openEMS} T.~Liebig, A.~Rennings, S.~Held, and D.~Erni, ``openEMS: A free and open source equivalent-circuit (EC) FDTD simulation platform supporting cylindrical coordinates suitable for the analysis of traveling wave MRI applications,'' \emph{Int. J. Numer. Model.}, vol.~26, no.~6, pp.~680--696, 2013.
\end{thebibliography}
\end{document}